\documentclass[pre,twocolumn,amsmath,amssymb,floatfix, showpacs]{revtex4}
\usepackage{ulem}

\usepackage{graphicx}% Include figure files
\usepackage{dcolumn}% Align table columns on decimal point
\usepackage{bm}        % Bold math
\usepackage{color}
\begin{document}
\title{{Large-scale anisotropy in stably stratified rotating flows}}
\author{R.~Marino$^{1,2}$, P.D.~Mininni$^{3}$, D.L.~Rosenberg$^4$, and  A.~Pouquet$^{5,6}$}
\affiliation{
$^{1}$Institute for Chemical-Physical Processes,  Rende (CS), 87036, ITALY.\\
$^{2}$Space Sciences Laboratory, University of California, Berkeley, CA 94720, USA. \\
$^{3}$Departamento de F\'\i sica, Facultad de Ciencias Exactas y Naturales, Universidad de Buenos Aires \& IFIBA, CO\-NI\-CET, Ciudad Universitaria, 1428 Buenos Aires, ARGENTINA.\\
$^{4}$National Center for Computational Sciences, Oak Ridge National Laboratory, P.O. Box 2008, Oak Ridge, TN 37831, USA.\\
$^{5}$Laboratory for Atmospheric and Space Physics, University of Colorado at Boulder, Boulder, CO 80309, USA.
$^{6}$National Center for Atmospheric Research, P.O.~Box 3000, Boulder, CO 80307, USA.
}
\date{\today}

\begin{abstract}
We present results from direct numerical simulations of the Boussinesq equations in the presence of rotation and/or stratification, both in the vertical direction. The runs are forced isotropically and randomly at small scales and have spatial resolutions of up to $1024^3$ grid points and Reynolds numbers of $\approx 1000$. We first show that  solutions with negative energy flux and inverse cascades develop in rotating turbulence, whether or not stratification is present. However, the purely stratified case is characterized instead by an early-time, highly anisotropic transfer to large scales with almost zero net isotropic energy flux. 
This is consistent with previous studies that observed the development of vertically sheared horizontal winds, although only at substantially later times. However, and unlike previous works, when sufficient scale separation is allowed between the forcing scale and the domain size, the total energy displays a perpendicular (horizontal) spectrum with power law behavior compatible with $\sim k_\perp^{-5/3}$, including in the absence of rotation. In this latter purely stratified case, such a spectrum is the result of a direct cascade of the energy contained in the large-scale horizontal wind, as is evidenced by a strong positive flux of energy in the parallel direction at all scales including the largest resolved scales.
\end{abstract}

      % Direct numerical simulations

\maketitle
\section{Introduction} 

{The atmosphere and the global ocean are both forced by solar radiation and tides, together with surface winds and bathymetry for the ocean. The forcing scale can vary from planetary scales ($\approx 10^4$ km) all the way down to $1$ km for ocean floor topography. Under large-scale quasi-geostrophic balance (a balance established between gravity, rotation and pressure gradients), the motions are quasi two-dimensional and the flow of energy is thought to be predominantly to the large scales. However, this poses several problems, one of which being the way energy  in the whole system is being transferred to and dissipated at small scales of the order of millimeters (see, e.g., \cite{deusebio_13} for a recent review, and references therein). This issue is important for the understanding of the large-scale spectrum of the atmosphere, and in the closing of the global overturning circulation of the ocean, as described recently for example in \cite{thurnherr_11, nikurashin_13}. With a rotation rate at mid latitudes of $\Omega \approx 10^{-4}$ s$^{-1}$, and Brunt-V\"ais\"al\"a frequencies varying from $10^{-3}$ s$^{-1}$ as in the Southern abyssal Ocean \cite{nikurashin_12} to ten times that in the atmosphere and stratosphere, there are a variety of high Reynolds number turbulent regimes that occur in geophysical flows. Therefore, mesoscales in the atmosphere are dominated by stratification, while the Southern abyssal Ocean has stratification and rotation of comparable magnitude (although stratification is still dominant). Of course, current supercomputers do not allow to study realistic values of the Reynolds number for such flows, but is should be  high enough in order to include the effect of small-scale turbulent eddies on large-scale dynamics \cite{bartello_13}, hence the need for large numerical resolution.}

{As a result of this variety of regimes and of the limitations in computer power,} in recent years progress has been made in understanding rotating stratified flows thanks to a combination of tools, including, e.g., wave turbulence approaches \cite{Cambon01, newell_01, Sagaut, nazar}, reduced equations based on asymptotic expansions in a small parameter \cite{embid_majda_98, Julien06, klein_rev_10, wingate_11}, experiments \cite{baroud_03, troy_05, praud06, vanBokhoven_09, bordes_12}, observations in the atmosphere and the ocean \cite{nastrom_06, li_09}, and direct numerical simulations \cite{Aluie11,Waite11,Kimura12,Almalkie12, bartello_13} (see also the reviews in \cite{riley_rev_00, staquet_rev_02}). However, some fundamental issues have not been clarified yet. For example: What is the distribution of energy among modes (the Fourier spectrum)? What is the role of anisotropy, and under what conditions does the system self-organize and develop large-scale structures?

In many turbulent flows this self-organization takes place through a so-called inverse cascade. In particular, in a forced flow and under certain conditions, when the forcing is applied at a rather small scale compared to the size of the domain, scales larger than the forcing can be excited in a self-similar process. When the process results in a power law spectrum at small wavenumbers, with constant negative flux of some quantity that is conserved in the ideal limit, the system is said to sustain an inverse cascade of that quantity. It is a well known feature of nonlinear dynamics that has been reviewed extensively \cite{Biferale11,Boffetta12, amrik_rev}, in particular in the context of two dimensional (2D) Navier-Stokes fluids \cite{Kraichnan80}, and that has also been
observed in astrophysical plasmas \cite{marino_08, marino_11, marino_12}.

Direct numerical simulations have shown that rotating flows in finite domains can develop inverse cascades of energy (see, e.g., \cite{smith_99, Sen12}), and recent experiments have observed the same phenomena \cite{Yarom13}. In the case of stratified flows, the situation is less clear. Conclusions from simulations are sometimes contradictory and depend on whether the flow also rotates or not. When rotation and stratification are comparable, inverse cascades have been observed \cite{bartello_95,Metais96,Kurien08,Pouquet2013}. In this case, helicity is also created in the flow \cite{hide_76, marino}, and can be responsible for the so-called anisotropic kinetic alpha effect instability \cite{frisch_AKA_87}, as well as for other large-scale instabilities \cite{tur_13,tur_13b}. For stratified flows with little or no rotation, it was noted in \cite{Smith02} that for very long times an inverse transfer develops, but with no clear evidence of a self-similar spectrum at large scales. This inverse transfer generates vertically sheared horizontal winds, that were also reported in \cite{Laval03}. However, statistical mechanics arguments used in \cite{Waite04,Waite06, herbert_14}, based on the conservation of energy and linearized potential vorticity, lead one to argue against the existence of an inverse cascade in purely stratified flows.

{Recently we showed that in the opposite limit, when stratification is weak compared with rotation, the inverse cascade that develops is faster than in the purely rotating case \cite{Marino13}. We have also provided evidence of the generation of a dual energy cascade in Boussinesq flows with rotation, when forcing is applied at intermediate scales \cite{Pouquet2013}. In these previous studies, our investigations of the purely stratified case were restricted to the study of the isotropic Fourier modes and did not show any occurrence of an inverse energy cascade.}

{By means of more detailed field decompositions, we present here results from high resolution numerical simulations of rotating and/or stratified flows (up to $1024^3$ spatial grid points), forced at small scales, to point out the existence of an inverse energy transfer even in Boussinesq flows without rotation. However, this transfer is highly anisotropic, and is different from the fluxes in inverse cascades reported in the rotating and stratified cases. In particular, we show that while solutions with negative energy flux and inverse cascades develop in rotating stratified flows, as observed in 
\cite{Marino13,Pouquet2013}, the purely stratified case is characterized instead by a highly anisotropic transfer with almost zero isotropic energy flux at large scales (i.e., no inverse cascade develops).} In this latter case, the perpendicular kinetic energy spectrum develops a peak at scales larger than the forcing scales, which is consistent with findings in previous works that reported the development of vertically sheared horizontal winds \cite{Smith02,Laval03}. However, and unlike previous works, when sufficient scale separation is allowed between the forcing scale and the domain size, the energy displays a perpendicular spectrum with power law behavior compatible with $\sim k_\perp^{-5/3}$. This spectrum is the result of a direct cascade of the energy contained in the large-scale horizontal velocity, and is accompanied by a positive flux of energy in the parallel direction at the largest scales. This result is interesting in the light of observations of a spectrum with similar scaling and with positive flux at intermediate scales in the atmosphere \cite{Nastrom84,Lindborg99,Heas12}, as it may provide another mechanism to generate such a power law (see \cite{Tung03,Tulloch06,Brethouwer07,Riley08,Teitelbaum12,Kimura12} for other explanations).

\section{Simulations of rotating and stratified flows}

We use a parallelized pseudo-spectral code \cite{Gomez05,Mininni11} to solve the Boussinesq equations for an incompressible and stratified fluid:
\begin{equation}
\frac{\partial {\mathbf u}}{\partial t} + {\mathbf u} \cdot \nabla {\mathbf u} - \nu \nabla^2 {\mathbf u} = -\nabla p - N \theta \hat{z} - f \hat{z} \times {\mathbf u} + {\bf F} ,
\label{eq:mom}
\end{equation}
\begin{equation}
\frac{\partial \theta}{\partial t} + {\mathbf u} \cdot \nabla \theta - \kappa \nabla^2 \theta = N {\mathbf u} \cdot \hat{z} ,
\label{eq:temp}
\end{equation}
\begin{equation}
\nabla \cdot {\bf u}=0 .
\end{equation}
Here, ${\mathbf u}$ is the velocity, $\theta$ the (potential) temperature fluctuations, and $p$ the pressure normalized by a unit mass density. The pressure is obtained self-consistently from the incompressibility condition. We consider a Prandtl number $Pr=\nu/\kappa=1$, with $\nu$ the kinematic viscosity and $\kappa$ the temperature diffusivity. In Eqs.~(\ref{eq:mom}) and (\ref{eq:temp}), $N$ is the  Brunt-V\"ais\"al\"a frequency, and $f=2\Omega$ with $\Omega$ the rotation frequency. The external mechanical force ${\bf F}$ is isotropic and generated randomly, applied in a shell of modes with wavenumbers $k_F$ (resulting in a forcing length scale $L_F=2 \pi/\textrm{min}\{k_F \}$). Although many studies of purely stratified flows use 2D forcing to mimic atmospheric conditions, we use isotropic three dimensional (3D) forcing instead as we will compare simulations with pure rotation ($N=0$) with simulations with pure stratification ($f=0$), and we want to let the system develop in each case the level of anisotropy that is self-consistent with the level of rotation and stratification that is externally imposed (see below). The equations are solved in a triple periodic domain of length $2 \pi$ with $n_p$ grid points in each direction, and evolved in time using a second order Runge-Kutta method.

\begin{table}
\caption{\label{table:runs} {Parameters used in the simulations: $n_p$ is the 
linear grid resolution, $k_F$ the forcing wavenumber, $\textrm{Ro}$ the Rossby 
number, $\textrm{Fr}$ the Froude number, $k_Z$ the Zeman scale, and $k_O$ the 
Ozmidov scale (see text for definitions). The Reynolds number in all the 
runs is $\approx 1000$.}}
\begin{ruledtabular}
\begin{tabular}{ccccccc}
  Run &         $n_p$ & $k_F$     & $\textrm{Ro}$ & $\textrm{Fr}$ & { $k_Z$} & {$k_O$}  \\
\hline
   I       &    1024  & $[40,41]$   &     0.08     &  0.04   & {440}    &   {880}  \\ % 5
   II      &    1024  & $[40,41]$   &   $\infty$ &  0.04   & {  --}    &   {512}  \\ % 7
   \hline
   III      &    512   & $[22,23]$   &     0.08      &  0.04     &  {210}   &   {420}  \\ % 31 
   IV      &    512   & $[22,23]$   &   $\infty$  &  0.04     &  { --}    &   {250}  \\ % 33
   V       &    512   & $[22,23]$   &   $\infty$  &  0.02     &  { --}    &   {490}  \\ % 35
   VI      &    512   & $[22,23]$   &   $\infty$  &  0.08     &  { --}    &   {120}  \\ % 36
   VII     &    512   & $[22,23]$   &   $\infty$  &  0.12     &  { --}    &   {80}    \\ % 37
   VIII    &    512   & $[22,23]$   &     0.08    & $\infty$  &  {240}   &   {  --}  \\ % 32
\end{tabular}     \end{ruledtabular}  \end{table}

The parameters of all the simulations are given in Table \ref{table:runs}. The simulations are characterized by dimensionless numbers based on a characteristic unit velocity $U=1$: the Reynolds number
\begin{equation}
\textrm{Re} = \frac{UL_F}{\nu} ,
\end{equation}
the Froude number
\begin{equation}
\textrm{Fr} = \frac{U}{N L_F} ,
\end{equation}
and the Rossby number
\begin{equation}
\textrm{Ro} = \frac{U}{f L_F} .
\end{equation}
Time in the simulations below is measured in units of the turnover time at the forcing scale, defined as $\tau_{NL} = L_F/U$.

{One can also define several characteristic length scales for these flows, such as the Zeman and Ozmidov scale at which isotropy recovers (when respectively under the sole influence of rotation or stratification),
\begin{equation}
\ell_Z= 2\pi (\epsilon/f^3)^{1/2}                       \ \ , \ \ \ell_{O}=2\pi (\epsilon/N^3)^{1/2} \ .
\end{equation}
Here $\ell_Z$ is the Zeman scale, $\ell_{O}$ the Ozmidov scale, and $\epsilon$ the rate of energy dissipation. The wavenumbers associated with these scales, respectively $k_Z$ and $k_{O}$ (with $k = 2\pi/\ell$), are given in Table \ref{table:runs} for the different runs. Under the hypothesis that isotropy has recovered at small scale, one can also define the Kolmogorov dissipation scale
\begin{equation}
\eta=2\pi (\epsilon/\nu^3)^{-1/4} \ ,
\end{equation}
with associated wavenumber $k_\eta = 2\pi/\eta$. In the $1024^3$ runs this wavenumber is $\approx 340$, while in the $512^3$ runs this wavenumber is $\approx 200$. Note that based on these wavenumbers, most of the simulations presented here are anisotropic at all scales (i.e., even at the smallest resolved scales), and there is a resonable scale separation between the different relevant physical scales in the system.}

We start the simulations from isotropic random initial conditions with a steep spectrum ($\sim k^{-4}$ followed by an exponential decay) from $k_F$ to the maximum wavenumber $k_{max}=n_p/3$. At $t=0$, no energy is present in any Fourier mode with $k<k_F$. The forcing is applied during the entire run, and the systems are evolved for at least 35 turnover times $\tau_{NL}$. During the run we monitor the evolution of the total energy to see whether there is a growth that may indicate the development of an inverse cascade, or the growth of large-scale structures. In practice, energy grows or decreases first during a short transient (less than $10 \tau_{NL}$), and then the systems evolve towards a quasi-stationary state with approximately constant energy, or with energy growing monotonically in time when an inverse cascade develops. To verify whether the small scales in the system have reached a turbulent steady state, we monitor as well the time evolution of the enstrophy (proportional to the energy dissipation rate).

In the absence of stratification, the inertial wave frequency is $\omega_{\bf k} = f k_{\parallel}/k$, with  $k=|{\bf k}|$ and with $k_{\parallel}$ referring to the direction of imposed rotation and/or stratification and $k_\perp$ denoting the plane perpendicular to this. In the purely stratified case, gravity waves satisfy 
$\omega_{\bf k} = N k_\perp/k$ with $k_\perp = (k_x^2 + k_y^2)^{1/2}$, and in the general case 
\begin{equation}
\omega_{\bf k}=k^{-1} \sqrt{N^2k_{\perp}^2+f^2k_{\parallel}^2} \ .
\label{eq:disp}
\end{equation}
Note that the slow modes (modes with zero frequency) in the purely rotating case are the modes with $k_{\parallel}=0$ (i.e., 2D modes), while in the purely stratified case, they are the modes with $k_{\perp}=0$. The slow zero eigenvalue modes in the general (rotating and stratified) case in a periodic box with no mean flow have a more complex expression, see e.g., \cite{bartello_95, herbert_14}. This further explains why we use 3D forcing, as using 2D forcing would favor direct injection of energy only in the slow modes in the purely rotating case, and in a combination of slow and fast (wave) modes in any other case.

In order to characterize the anisotropy that develops in the flow, we use different definitions for the energy spectrum, as well as for anisotropic fluxes. We  start from the axisymmetric kinetic energy spectrum defined as \cite{Godeferd03,Mininni12}:
\begin{eqnarray}
{}&& e(k_{\perp},k_{\parallel})=
    \frac{1}{2} \sum_{\substack{
          k_{\perp}\le |{\bf k}\times \hat {\bf z}| < k_{\perp}+1 \\
          k_{\parallel}\le {\bf k} \cdot \hat{z} < k_{\parallel}+1}} |\tilde{\bf u}({\bf k})|^2 
          \nonumber \\
{}&& = \frac{1}{2} \int |\tilde{\bf u}({\bf k})|^2 k \sin \Theta \, 
          d \phi  = e(k, \Theta) \ ,
\label{etheta}
\end{eqnarray}
with the tilde denoting Fourier transform, $\Theta = \tan^{-1}(k_\perp/k_\parallel)$ the co-latitude in Fourier space with respect to the vertical axis with unit vector $\hat {z}$, and $\phi$ the longitude with respect to the $x$ axis. The sum corresponds to the discrete Fourier version used in the numerical simulations, while the integral is the definition in spherical coordinates in a continuous space.

From the axisymmetric kinetic energy spectrum one can define the so-called reduced spectra $E$ as a function of $k_{\perp}$, $k_{\parallel}$, and $k$ as:
\begin{eqnarray}
E(k_{\perp}) &=& \int e(k_{\perp},k_{\parallel}) d k_{\parallel} \ , \\ 
E(k_{\parallel}) &=& \int e(k_{\perp},k_{\parallel}) d k_{\perp} \ , \\ 
E(k) &=& \int e(k, \Theta) k \, d \Theta \ .
\label{reduced}
\end{eqnarray}
Note these spectra correspond to integration of the velocity correlation tensor in Fourier space respectively over cylinders, planes, and spheres. All the reduced spectra have the physical dimension of an energy density, i.e., summed over wavenumber they all yield, through Parseval's theorem, the kinetic energy $E_V=\langle |{\bf u}|^2\rangle /2$. However, the axisymmetric spectrum $e(k,\Theta)$ has units of energy per square wavenumber, and as a result an isotropic flow with spectrum $E(k) \sim k^{-\alpha}$ has axisymmetric spectrum $e(k, \Theta) \sim k^{-\alpha-1}$. Following the same procedure, spectra for the potential energy $E_P$ and for the total energy $E_T=E_V+E_P$ can be defined.

\begin{figure}
\includegraphics[width=8.6cm]{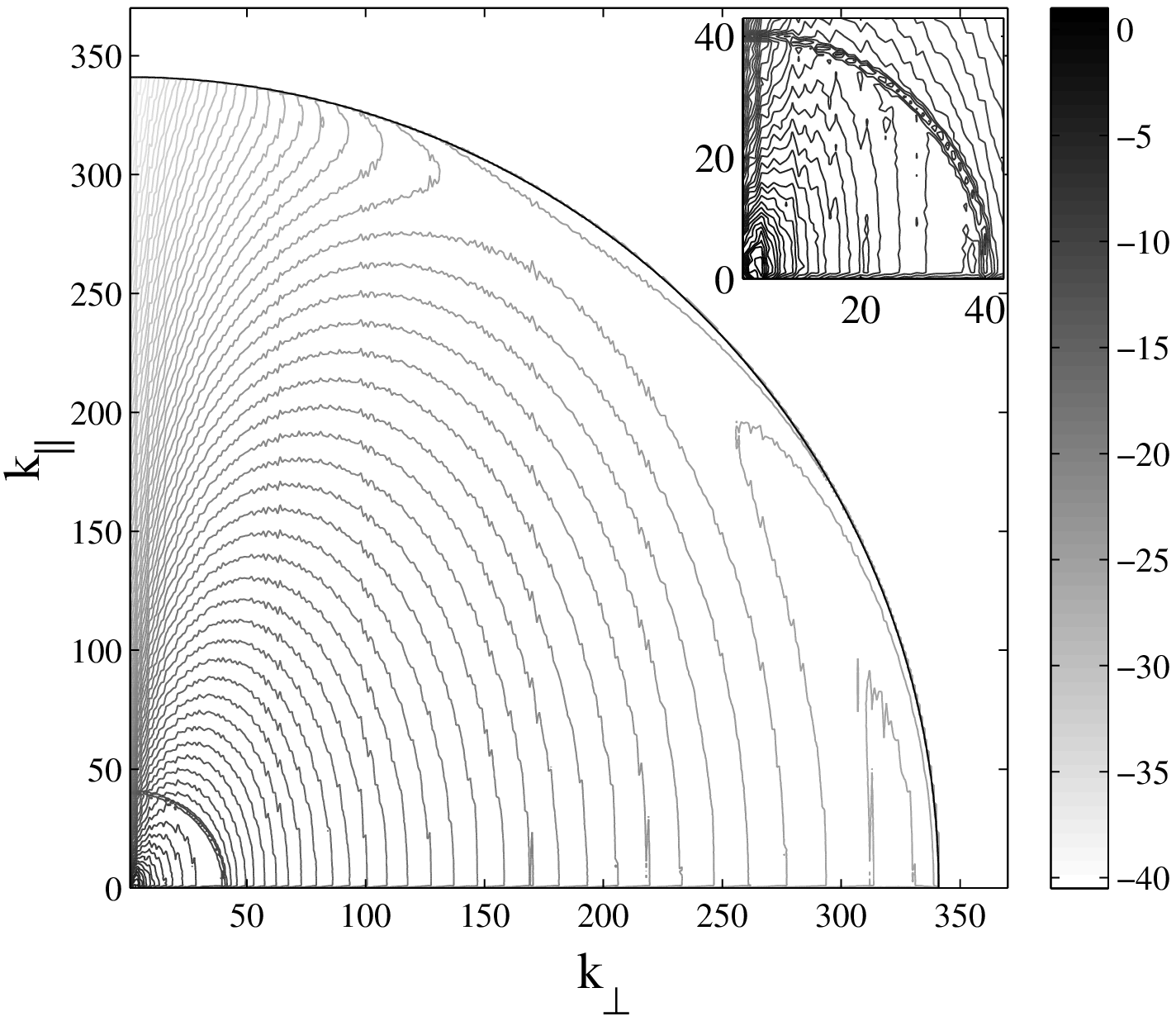}
\includegraphics[width=8.6cm]{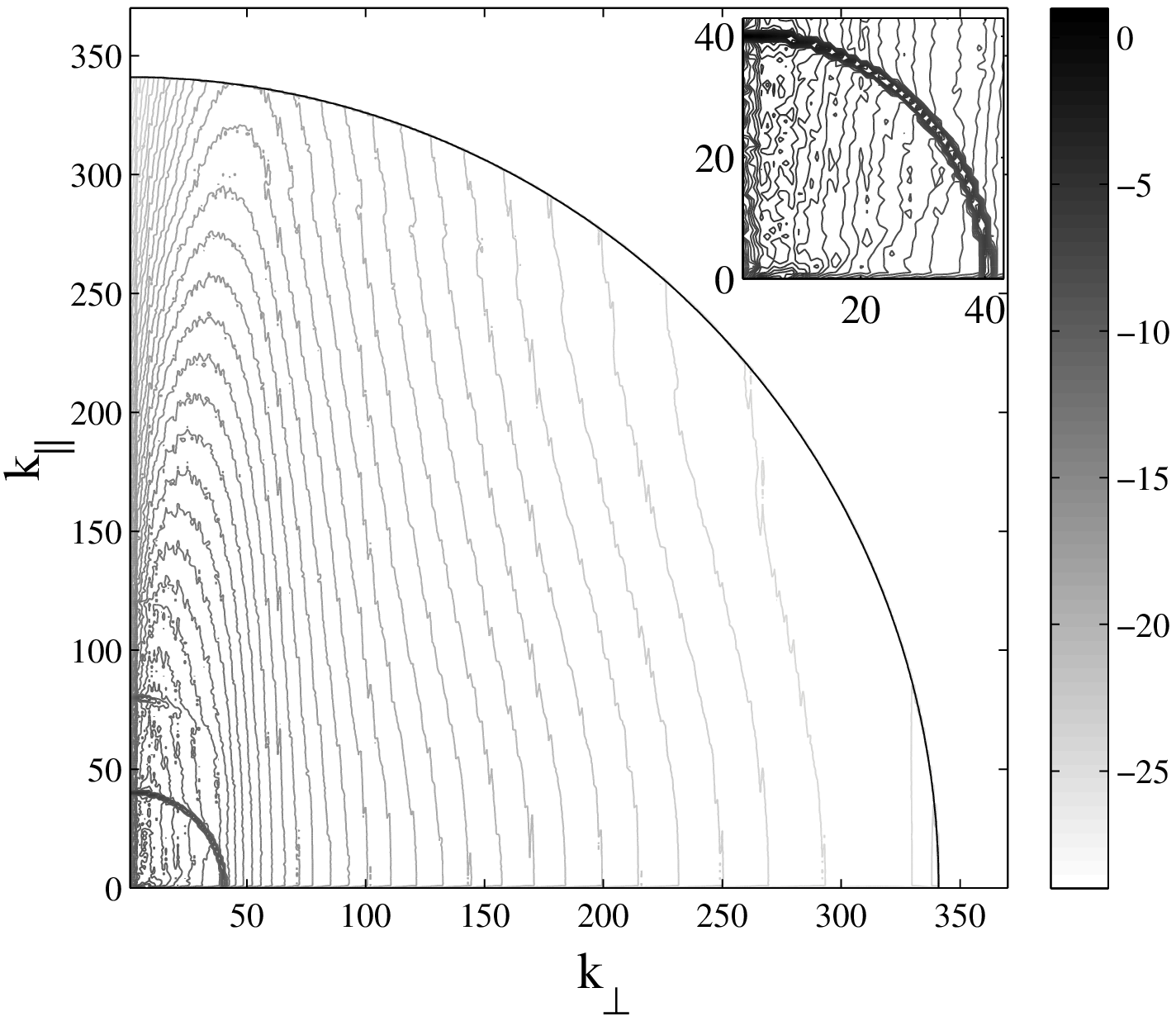}
\caption{{Contour plots of the logarithm of the axisymmetric kinetic energy spectrum 
$e(k_\perp,k_\parallel)$ for a rotating and stratified flow with $N/f =2$ (above, 
run I) and for a purely stratified flow with $\textrm{Fr}=0.04$ (below, run II), 
averaged in time over $20 < t/\tau_{NL} < 30$}. Note that the spherical shell 
of externally forced modes is clearly visible at $k \in [40,41]$. The two insets 
show  details of the axisymmetric spectra at wavenumbers smaller than 
or equal to the forcing wavenumbers.}
\label{fig:2Dspectrum} \end{figure}

All these spectra give information on the  anisotropy of the flow in Fourier space, but they do not provide information on how energy is transferred in that space. To study the spectral transfer, we thus also consider anisotropic fluxes. We start from an axisymmetric transfer function  for the total energy:
\begin{eqnarray} 
{}&& t(k_{\perp},k_{\parallel}) = \int \left[ \tilde{\bf u}({\bf k}) \cdot 
          \widetilde{\left( {\bf u} \cdot \nabla {\bf u} \right)}^*_{\bf k} 
          \right. \nonumber \\
{}&& + \left. \tilde{\theta}({\bf k}) \widetilde{\left( {\bf u} \cdot 
          \nabla \theta \right)}^*_{\bf k} \right] k \sin \Theta \, 
          d \phi + \textrm{c.c.} \ ,
\end{eqnarray}
with $^*$ and $\textrm{c.c.}$ denoting complex conjugate. This axisymmetric transfer function can be written in spherical coordinates as well, and will be denoted as $t(k, \Theta)$.

\begin{figure}
\includegraphics[width=8.6cm]{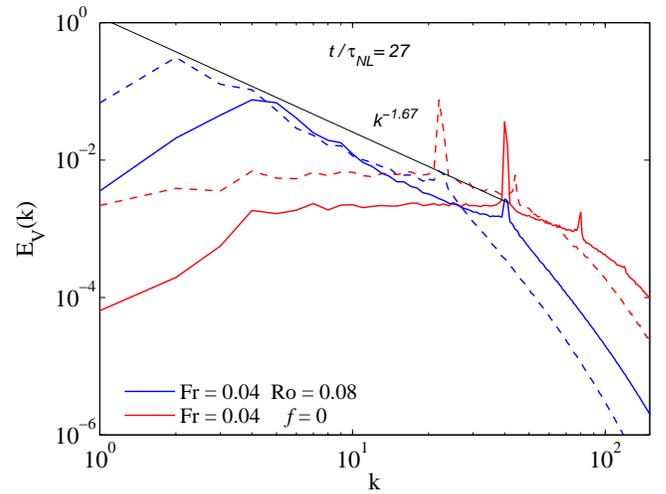}
\caption{{\it (Color online)} Isotropic kinetic energy spectrum $E_V(k)$ at $t=27 \tau_{NL}$ in simulations with the same Reynolds number, with $n_p=512$ linear grid points and forced at $k_F=[22,23]$ (dashed), and with $n_p=1024$ and $k_F=[40,41]$ (solid). 
Two cases are shown: a purely stratified case with $\textrm{Fr}=0.04$ (light or red curves), and a rotating stratified case with $N/f=2$ (dark or blue curves). A Kolmogorov slope is indicated as a reference.}
\label{fig:isotropic}
\end{figure}

Following the same procedure as with the energy, we can integrate over cylinders, planes and spheres in Fourier space to obtain
\begin{eqnarray}
T(k_{\perp}) &=& \int t(k_{\perp},k_{\parallel}) d k_{\parallel} \ ,  \\ 
T(k_{\parallel}) &=& \int t(k_{\perp},k_{\parallel}) d k_{\perp} \ ,  \\ 
T(k) &=& \int t(k, \Theta) k \, d \Theta \ .
\end{eqnarray}
We can finally define perpendicular, parallel and isotropic total energy fluxes respectively as
\begin{eqnarray}
\Pi_T(k_{\perp}) &=& -\int_0^{k_{\perp}} T(k_{\perp}') \, d k_{\perp}' 
          \ ,    \label{eq:flux1}  \\
\Pi_T(k_{\parallel}) &=& -\int_0^{k_{\parallel}} T(k_{\parallel}') \, 
          d k_{\parallel}' \ ,   \label{eq:flux2}  \\
\Pi_T(k) &=& -\int_0^k T(k') \, d k \ . \label{eq:flux3} 
\end{eqnarray}
These fluxes represent the total energy that goes across a cylinder, a plane or a sphere in Fourier space per unit of time, and should not be interpreted as partial fluxes (i.e., the three fluxes satisfy the condition that when integrated over all wavenumbers, they are equal to zero, as necessary to satisfy  total energy  conservation); for more details on anisotropic fluxes, see \cite{Mininni11b}.

\section{Results}

\begin{figure}
\includegraphics[width=8.6cm]{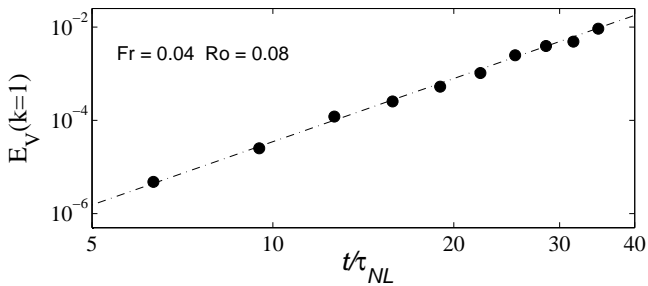}
\caption{{Evolution of the kinetic energy in the isotropic shell in Fourier space with $k=1$, as a function of time in log-log scale for Run I with $n_p=1024$ (see Table \ref{table:runs}). After an initial transient, the kinetic energy grows steadily with time, following a power law that is well approximated by $E_V(k=1,t)\sim t^{3.7}$. Accumulation of most of the energy in the $k=1$ shell does not take place in this simulation until much later times.}}
\label{fig:K_1} \end{figure}

\begin{figure*}
\includegraphics[width=8.6cm]{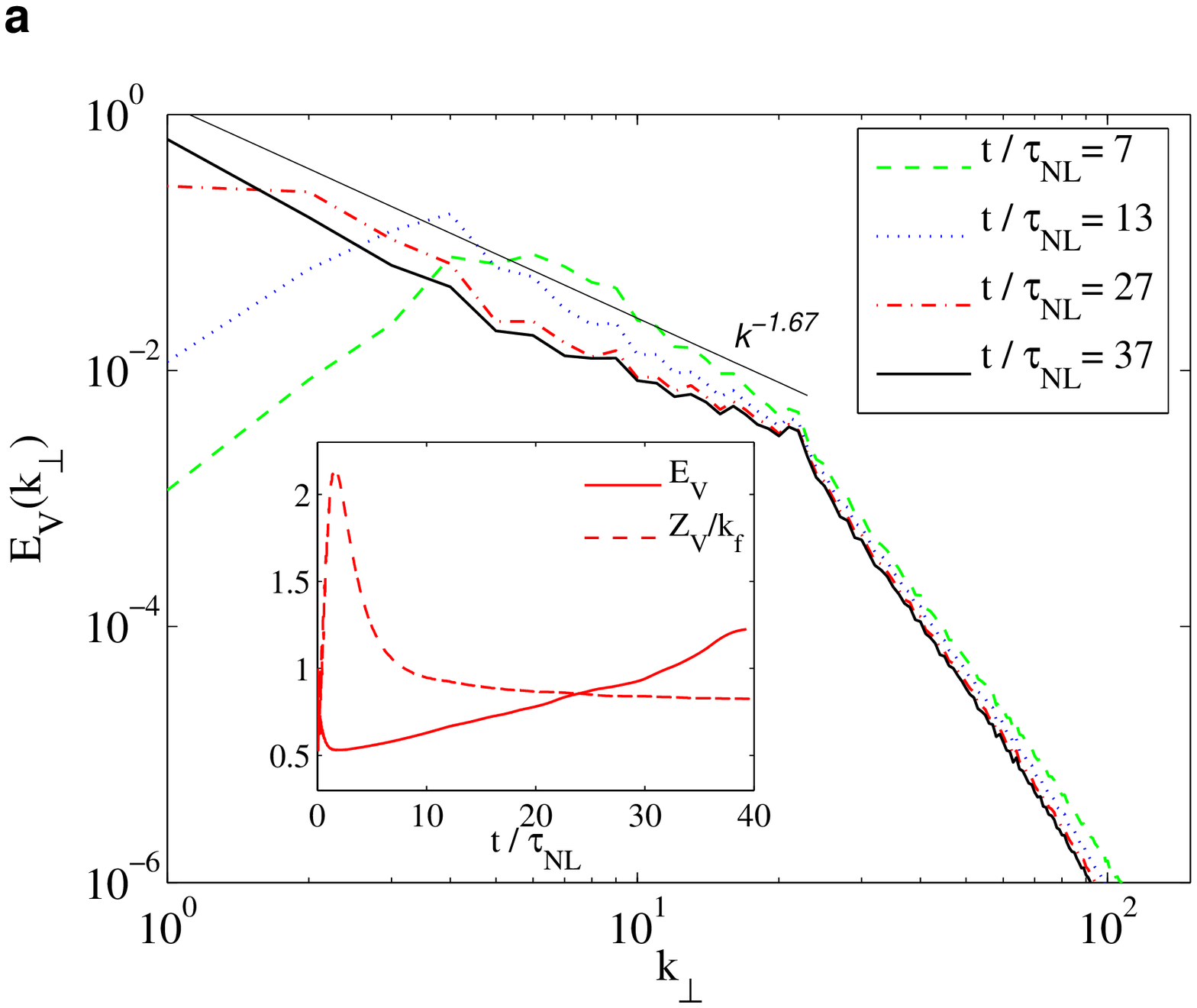}
\includegraphics[width=8.6cm]{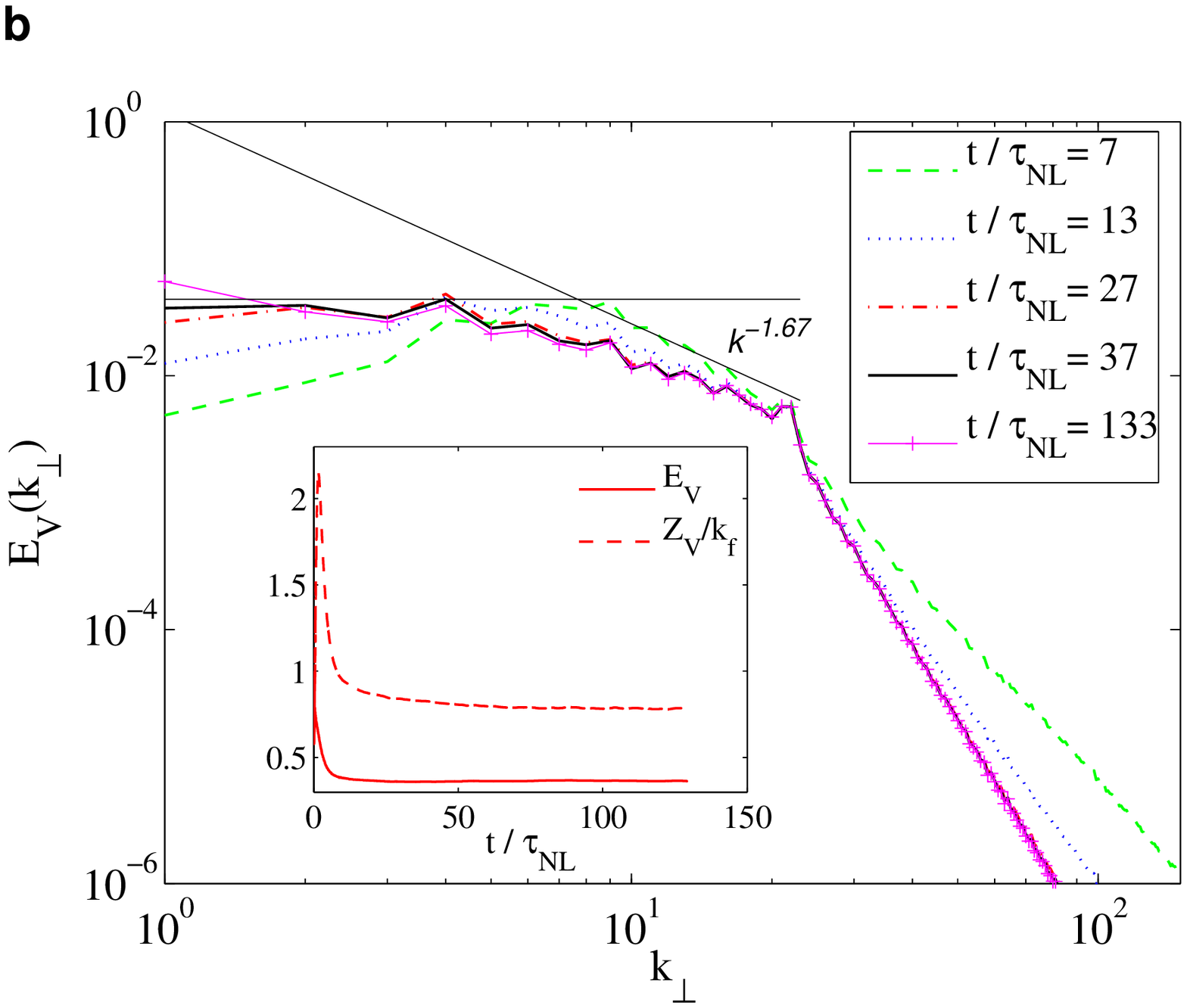}
\includegraphics[width=8.6cm]{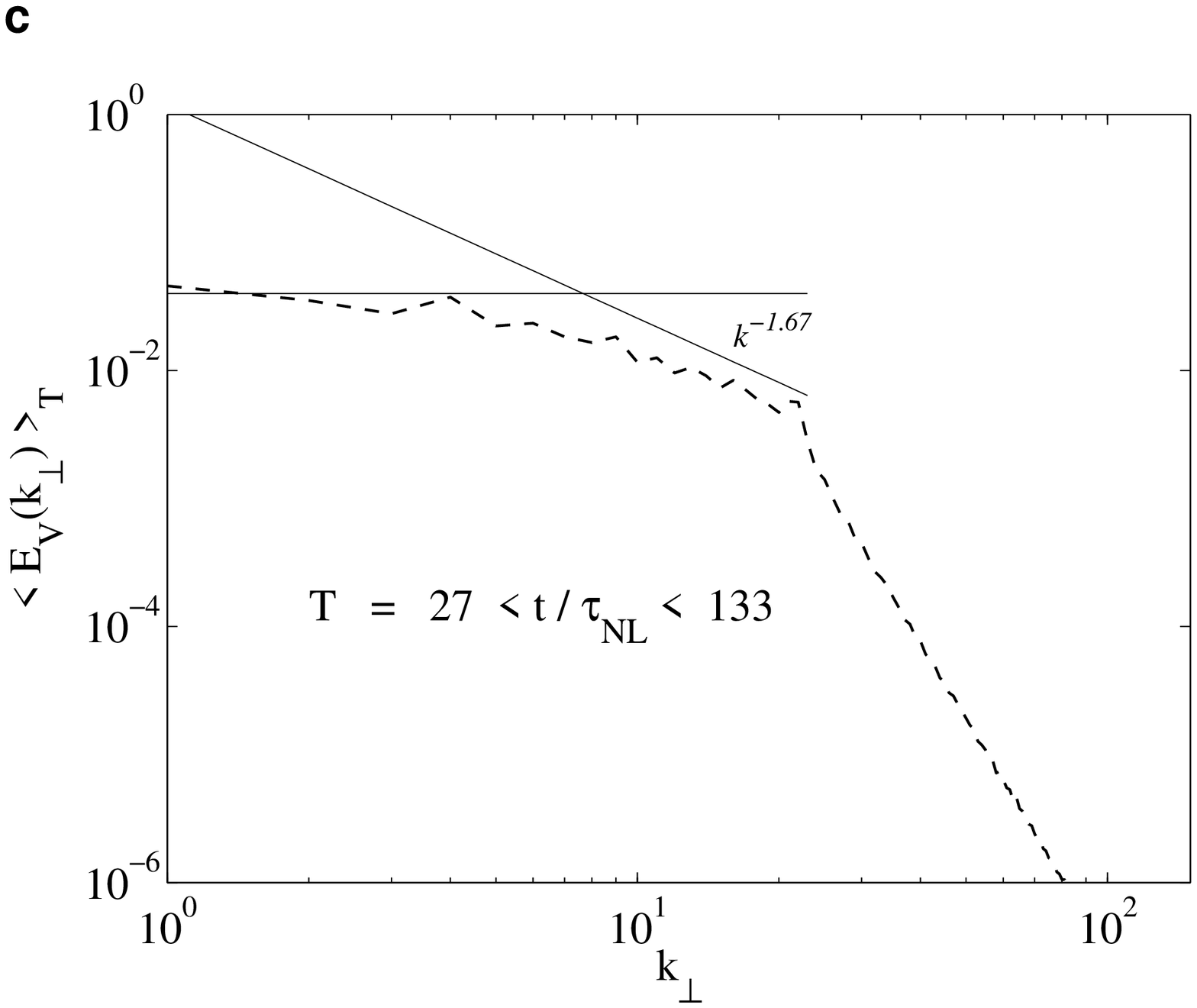}
\caption{{{\it (Color online)} Perpendicular kinetic energy spectra $E_{V}(k_\perp)$ 
at different times, for a simulation with $N/f=2$ (run III, panel a), and for a 
purely stratified flow (run IV, panel b), both with $\textrm{Fr}=0.04$. A 
Kolmogorov slope and a flat line are indicated as references. The insets in panel 
(a) and (b) show the time evolution of the kinetic energy and of the (normalized) 
kinetic enstrophy in the same runs (solid and dashed lines respectively).
Note the saturation of the kinetic enstrophy (i.e., of the power at the small 
scales) in both cases, and of kinetic energy only in the purely stratified case.
Panel (c) shows a time average over $\approx 100 \tau_{NL}$ of the the perpendicular 
kinetic energy spectrum for the purely stratified run of panel (b) (run IV).}
}
\label{fig:perpendicular} 
\end{figure*}

\subsection{Axisymmetric spectrum}

We start with a comparison of two configurations, one for a rotating and stratified flow, and the other one for a purely stratified flow. Figure \ref{fig:2Dspectrum} shows the axisymmetric kinetic energy spectrum $e(k_\perp,k_\parallel)$ in two $1024^3$ simulations: the first with $N/f=2$ (i.e., rotating and stratified), and the second with $f=0$ (only stratified). {The spectra are averaged in time over the range $20 < t/\tau_{NL} < 30$, and correspond to runs I and II in Table \ref{table:runs}. No significant differences are observed when instantaneous axisymmetric spectra (i.e., without averaging in time) are considered.
The modes that are forced are visible as dark isotropic contour levels (circles) with radius $k_F\approx 40$, and the flows are anisotropic at all other scales as is indicated by the shape of the remaining contours. Most of the energy concentrates near the axis $k_\perp = 0$ in the purely stratified case, and close to the line $k_\parallel = 2k_\perp$ in the run with $N/f=2$.} A run with pure rotation has most of the energy near the axis with $k_\parallel = 0$ (not shown). For small scales (i.e., for wavenumbers $k>k_F$) the development of these anisotropies is well known (see, e.g., \cite{Cambon01,Liechtenstein05}), and leads to the formation of layers and pancake-like structures in the stratified case, and to the formation of columnar structures in the rotating case. In the rotating and stratified case, the slope of the line that concentrates most of the energy in spectral space (in this case, $k_\parallel = 2k_\perp$) is associated with the development of layers in real space that are tilted with respect to the horizontal plane, and also with the growth of the vertical integral scale of the flow as rotation is increased \cite{Waite06}. The process underlying the formation of these structures in the general case is an anisotropic energy transfer resulting from resonant triads, that tends to move energy towards modes with zero frequency \cite{Waleffe93}, although strict wave resonance (without broadening) does not occur for $1/2 \le N/f \le 2$ as shown in \cite{Smith02} (see also \cite{Marino13}).

% 5
\begin{figure}
\includegraphics[width=8.6cm]{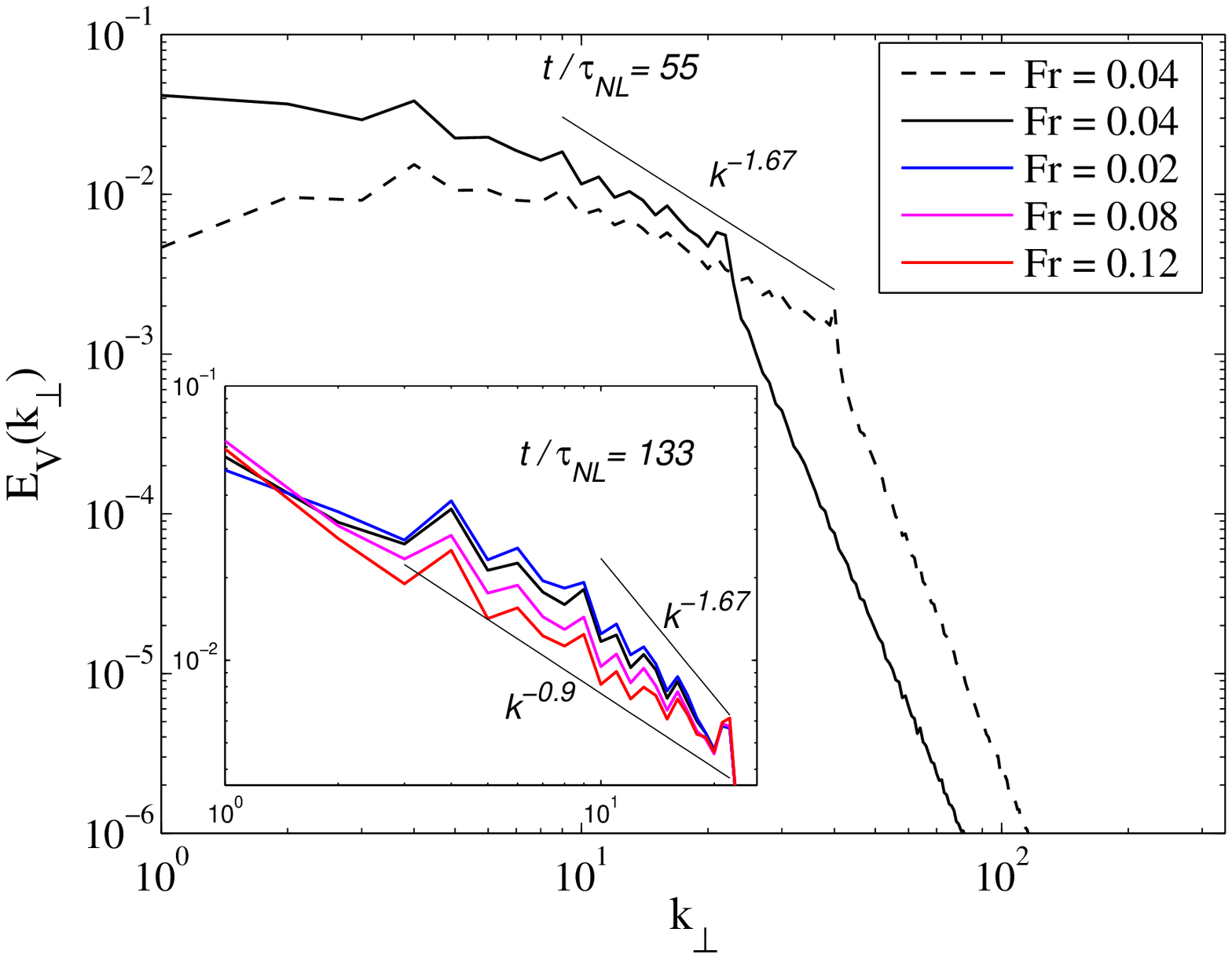}
\caption{{\it (Color online)} Perpendicular kinetic energy spectra $E_V(k_\perp)$ at $t=55\tau_{NL}$ for two purely stratified flows (runs II and IV) using grids with $n_p=512$ (solid) and $n_p=1024$ (dashed) points, with the same Froude and Reynolds numbers.
A Kolmogorov slope is given as a reference. The inset shows  details of the spectra for small wavenumbers at later times, for several runs with $\textrm{Re} \approx 1000$ (runs IV to VII), without rotation and with different Froude numbers. The labels in the top right box correspond to the curves in this inset.} 
\label{fig:multi} \end{figure}

Our focus here is not on small scale anisotropy, but rather on the anisotropy that develops at scales larger than the forcing. Note that for $k<k_F$ the spectra are still anisotropic, and the angular distribution  resembles what happens at $k>k_F$. In the purely stratified case, most of the energy is in modes with $k_\perp \approx 0$ (although, as will be shown later, energy is not concentrated in the modes with the smallest available perpendicular wavenumber). In the run with $N/f=2$, energy is more concentrated near the smallest wavenumber (i.e., at the largest scales available in the system), and unlike the purely stratified case, the modes with small wavenumber have more energy than the forced wavenumbers.

\subsection{Reduced spectra}

The difference in the energy contained at large scales in these systems can be also observed in the isotropic instantaneous spectra at late times, once the inverse cascade for $N/f=2$ is well developed. These spectra are shown in Fig.~\ref{fig:isotropic} for the two $1024^3$ simulations forced at $k_F \in [40,41]$, and also for two $512^3$ simulations forced at $k_F \in [22,23]$. While the rotating  stratified flows develop a power law spectrum at large scales (indicative of an inverse cascade), the purely stratified flows have a flat energy spectrum for $k<k_F$. In the purely stratified case, for $k>k_F$, the secondary peaks in the spectrum at $k\approx 45$ and $k\approx 81$ are likely due to harmonics of the forcing (centered at $k_F\approx 22.5$ in the $512^3$ run, and at $k_F\approx 40.5$ for the $1024^3$ run). The stronger prominence of harmonic frequencies in the purely stratified case may be due to the facts that 3D forcing excites stronger gravity waves, and that at these moderate Reynolds numbers the small-scale spectrum in the stratified case is also flatter and therefore energy in small scales is more prominent \cite{Kimura12}, a fact attributed to the accumulation of layers in the vertical direction.

{In the presence of an inverse cascade, energy grows steadily at the smallest wavenumbers. Since in these simulations no friction is used to remove the energy at large scales, all the analysis is stopped before accumulation (or ``condensation'') of energy at the gravest mode takes place, as this can result in a change in the slope of the spectrum. As an illustration, in Fig.~\ref{fig:K_1} we show the spectral energy at $k=1$ as a function of time for Run I (with $N/f=2$ and $n_p=1024$). Note the growth of the kinetic energy  in the gravest mode as a power law of time, and also note that at $t \approx 35 \tau_{NL}$ the energy $E_V(k=1)$ is still smaller than the energy in $k=2$ (see the solid dark curve in Fig.~\ref{fig:isotropic}), and also smaller than the total energy in the system.}

As is indicated by the axisymmetric spectrum in Fig.~\ref{fig:2Dspectrum}, the 
spectral distribution of the energy is strongly anisotropic, and the isotropic 
spectrum is insufficient to characterize the energy distribution at large scales. 
Figure \ref{fig:perpendicular} shows the reduced perpendicular kinetic energy 
spectrum $E(k_\perp)$ for two simulations as a function of time. In the rotating  
stratified flow with $N/f=2$ (run III), the energy spectrum grows at scales 
larger than the forcing scale, developing a power-law compatible with 
$\sim k_\perp^{-5/3}$ scaling. The peak of the spectrum moves towards larger 
scales (smaller wavenumbers) as time evolves, and the energy at $k_\perp=1$ grows 
steadily in time. After $t \approx 35 \tau_{NL}$, the perpendicular kinetic 
energy spectrum already peaks at $k_\perp=1$, and the analysis is stopped. At 
this time, most of the kinetic energy is concentrated in the largest available 
perpendicular scale in the system (although not in the largest isotropic scale, 
as explained above).

The kinetic enstrophy $Z=\left< \omega^2 \right>/2$ remains approximately constant 
in time after a transient, indicating the small scales in the flow have reached a 
steady state. A similar behavior is observed in the spectrum of the potential 
energy, and in the time evolution of the potential enstrophy. All these features 
suggest the development of an inverse cascade in the flow, which will be confirmed 
later by studying the energy fluxes.

% 6
\begin{figure}
\includegraphics[width=8.6cm]{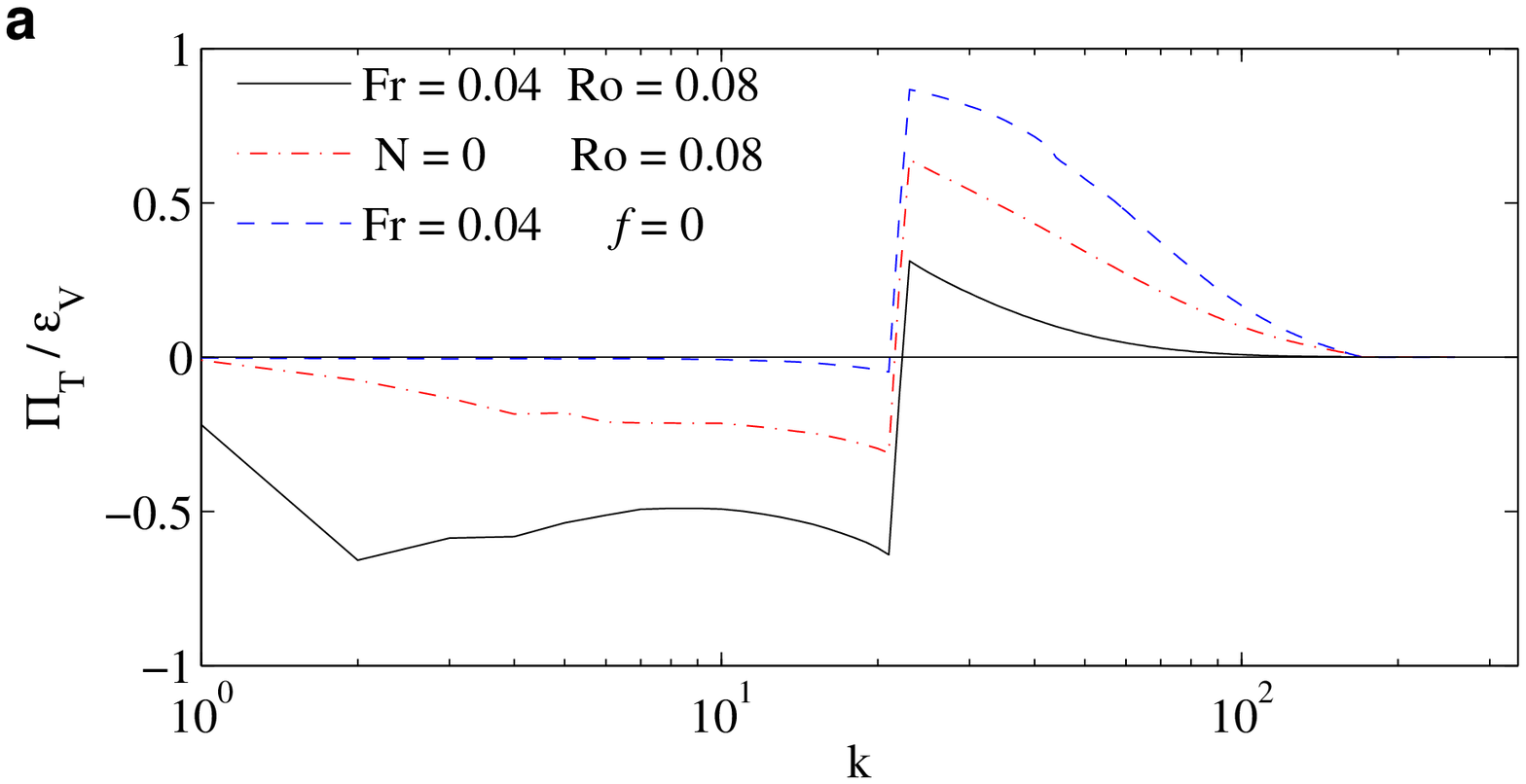}
\includegraphics[width=8.6cm]{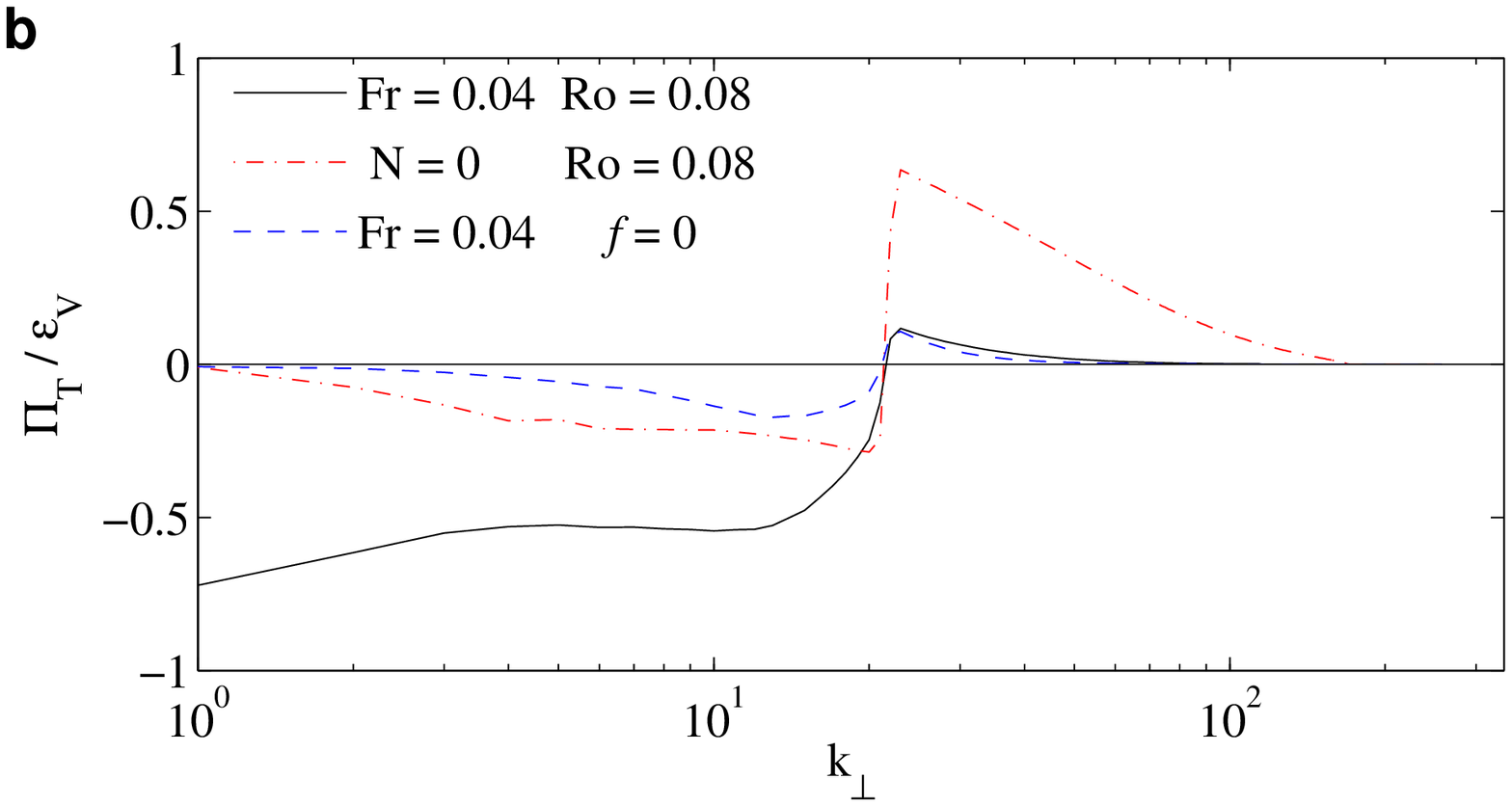}
\includegraphics[width=8.6cm]{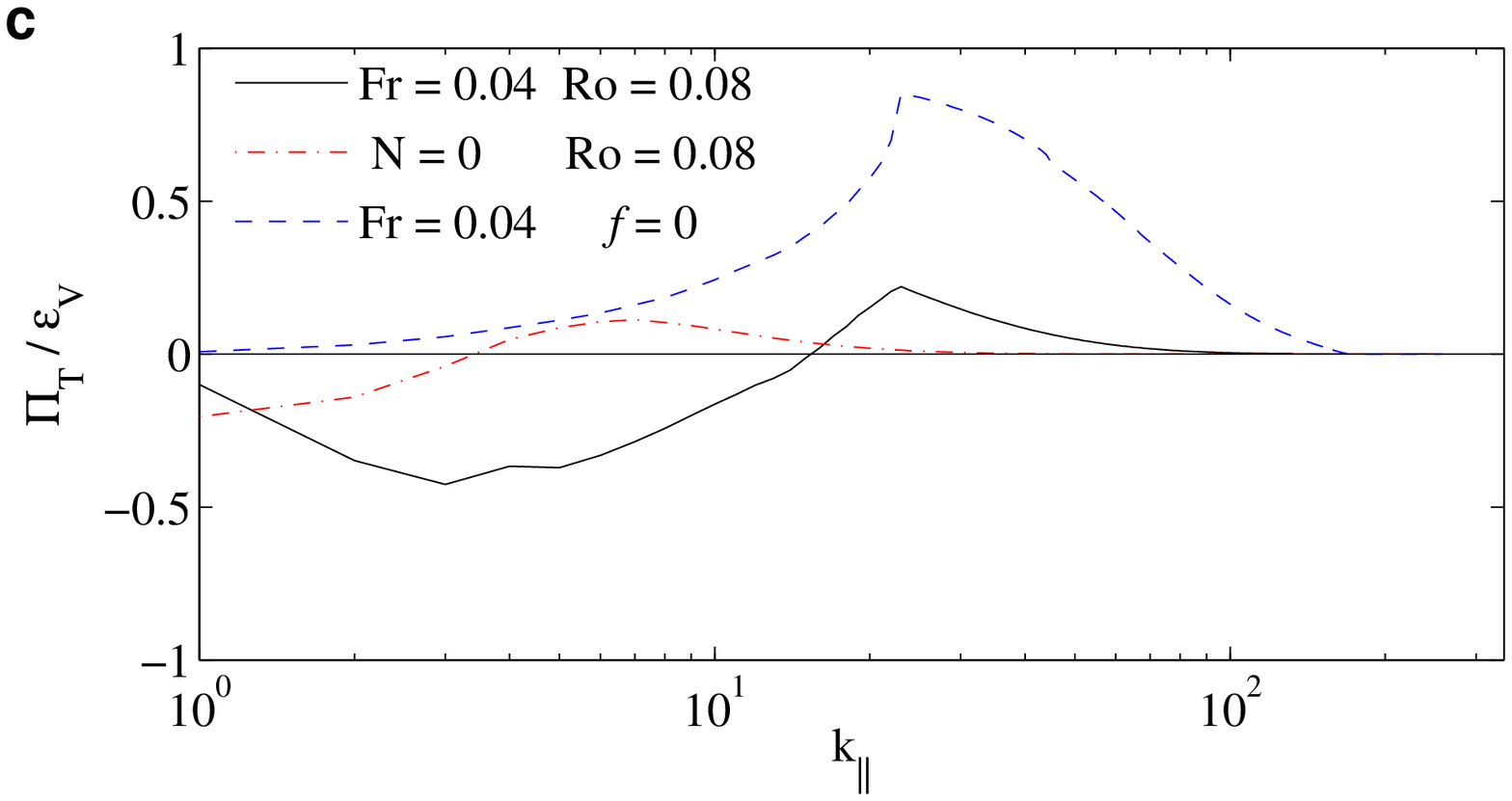}
\caption{
{\it (Color online)} (a) Isotropic, (b) perpendicular, and (c) parallel total 
energy fluxes normalized by the energy input $\epsilon_V=\left< {\bf u}\cdot {\bf F}\right>$ 
in a run with $N/f=2$ (solid, black line), a run with pure rotation (dash-dotted, 
red line), and a run with pure stratification (dashed, blue line).
}
\label{fig:flux} \end{figure}

The purely stratified case is different (run IV, Fig.~\ref{fig:perpendicular}b). 
The perpendicular kinetic energy spectrum $E(k_\perp)$ 
develops rapidly a narrow range with $\sim k_\perp^{-5/3}$ 
between $k_\perp \approx 4$ and $k_F$ (already visible at $t = 7 \tau_{NL}$, 
albeit in a narrower range of wavenumbers). This range of scales is 
followed at larger scales by a flat spectrum, as observed also in the 
isotropic spectrum $E(k)$ in Fig.~\ref{fig:isotropic}. {Evidence of flat 
spectra at large scale was already provided in simulations of stably 
stratified turbulence with an applied 2D forcing \citep{herring_89} 
(it must be pointed out that in that case, the absence of the vertical 
component of the forcing does not prevent the horizontal velocity field 
from developing a vertical energy flux in spectral space).} In Run IV, 
even for very long times ($t = 133 \tau_{NL}$) the flat spectrum at 
large scales persists, and the energy at $k_\perp =1$ does not grow 
significantly. The kinetic enstrophy $Z$ remains approximately 
constant in time after a transient (as in the run with $N/f=2$), but in 
this case the kinetic energy also remains approximately constant 
(unlike the rotating and stratified case).

{In order to demonstrate the persistence of the large-scale 
behavior of $E(k_\perp)$ in the purely stratified case, we averaged the 
spectrum over roughly one hundred turn-over times, i.e., over the 
interval $27 <  t / \tau_{NL} < 133$, as shown in 
Fig.~\ref{fig:perpendicular}c. A transition between a flat spectrum 
to a steeper spectrum can still be observed at $k_\perp \approx 4$ 
in the time-averaged reduced perpendicular spectrum.}

The differences in the run with $f=0$ between the isotropic and perpendicular kinetic energy spectra are associated with the fact that the axisymmetric spectrum is highly anisotropic, with most of the energy concentrated in modes close to the axis with $k_\perp=0$. As a result, integration of $e(k,\Theta)$ over circles or over planes with constant $k_\perp$ yields different reduced spectra. Is the reduced perpendicular spectrum an indication of an inverse cascade in the purely stratified case, even though it does not peak at the largest available perpendicular scale? It is interesting to point out that a spectrum peaking at intermediate wavenumbers was observed before in numerical simulations \cite{Smith02}. However, in the simulations in \cite{Smith02} no power law was found between the peak at intermediate wavenumbers and $k_F$, probably because the simulations were done at lower resolution and with less scale separation at large scales.

% 7
\begin{figure}
\includegraphics[width=8.6cm]{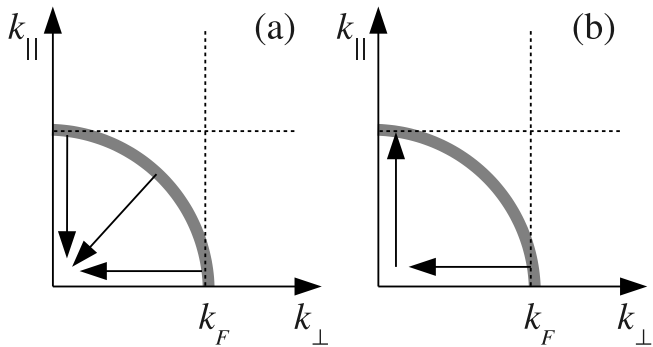}
\caption{
Schematic representation of the total energy  flux in Fourier space, defined in Eqs.~(\ref{eq:flux1}), (\ref{eq:flux2}), and (\ref{eq:flux3}), for (a) rotating (or rotating and stratified) flows, and (b) the purely stratified case. The gray circle indicates the shell of forced modes, and the dashed lines indicate planes across which the fluxes $\Pi_T(k_\perp)$ (represented by horizontal arrows)  and $\Pi_T(k_\parallel)$  (represented by vertical arrows) are computed. 
  The diagonal arrow represents the 
flux in terms of isotropic wavenumber, $\Pi_T(k)$ 
which, in case (b), is  negligible with two projections (on the $k_\perp$ and $k_\parallel$ directions) that are almost equal but of opposite sign. Note however that 
  the length of the arrows is arbitrary, and does not 
  denote the strength of each flux.
  }\label{fig:diagram} \end{figure}

To find the origin of this spectrum, we first study whether the peak of $E(k_\perp)$ 
at $k\approx 4$ in the stratified case depends on the scale separation or on the 
Froude number. Figure \ref{fig:multi} shows a run with $n_p=512$ forced at $k_F 
\in [22,23]$ (run IV), and a run with $n_p=1024$ forced at $k_F \in [40,41]$ 
(run II), both with $\textrm{Fr}=0.04$, $\textrm{Re} \approx 1000$ and with $f=0$. 
{This value of the Reynolds number is somewhat constrained by the choice of 
a small scale forcing but is still high enough for the flow to develop a power 
law turbulent spectrum}. 
In spite of the change in $k_F$, both runs show a peak at $k_\perp \approx 4$  
at the same time in units of turn-over time, with an approximately flat spectrum 
at smaller wavenumbers. 
The inset in Fig.~\ref{fig:multi} shows multiple $n_p=512$ runs (runs IV to VII) 
all forced at $k_F \in [22,23]$, all with the same Reynolds number  but now with different Froude numbers varying from $0.04$ to $0.12$ (see inset). 
As the effect of stratification becomes weaker (larger Froude number), the -5/3 range weakens as well and becomes shallower although, interestingly, a peak at $k_\perp \approx 1$ develops for the largest $\textrm{Fr}$.  
This could be attributed to the fact that, with a flatter spectrum, it is faster for the energy  to be available to larger scales.

\subsection{Fluxes and energy transfer}

We consider now the total energy fluxes defined in the previous section in 
Eqs.~(\ref{eq:flux1})-(\ref{eq:flux3}). The fluxes are displayed in Fig.~\ref{fig:flux} 
as a function of isotropic, perpendicular, and parallel wave numbers for three runs 
with $n_p=512$: one run with $\textrm{Fr}=0.04$ and $\textrm{Ro}=0.08$ (rotating 
and stratified, with $N/f=2$, run III), one run with $\textrm{Ro}=0.08$ and $N=0$ 
(purely rotating, run VIII), and one run with $\textrm{Fr}=0.04$ and $f=0$ (purely 
stratified, run IV). For $k>k_F$, the fluxes are positive in all the runs, 
indicating a direct cascade of total energy in the three cases. {To smooth 
out local temporal fluctuations of the energy transfer, fluxes are averaged 
over the range $20 < t/\tau_{NL} < 30$. This choice for the time average appears 
to be consistent with the quasi-stationarity of the inverse cascade as suggested 
in \citep{kraichnan_67}}. 

Although the scale separation between the minimum and forcing wavenumbers is insufficient to observe a range with constant fluxes, the sign of the flux is a clear indication of the direction of the energy transfer.

At scales larger than the forcing scale, or for $k<k_F$, the purely rotating run (dash-dotted red curves in Fig.~\ref{fig:flux}) has negative $\Pi_T(k)$ and $\Pi_T(k_\perp)$ fluxes. Note also that for that same run, $\Pi_T(k_{\parallel})$ is smaller and changing sign, although negative values do prevail for the smallest wavenumbers. Moreover, in $\Pi_T(k)$ and $\Pi_T(k_\perp)$, a range of wavenumbers can be identified at large scales for which the fluxes remain approximately constant. This is consistent with what was observed in previous studies (see, e.g., \cite{Smith02, Sen12}): in the purely rotating case, energy at large scales goes towards 2D modes (modes with $k_\parallel=0$), modes which  then undergo an inverse cascade in the 2D plane towards the largest scales in the system.

In the rotating and stratified run with $N/f=2$ (solid, black line), a similar behavior is observed, but the inverse fluxes are larger in absolute value (i.e., the inverse cascade is faster) than in the purely rotating case, while the direct fluxes are smaller. This is studied in more detail in  \cite{Marino13}; it is associated with the fact that for $N/f=2$ there are no resonant triads \cite{Smith02} and energy can be transferred more efficiently towards the 2D modes. For larger values of $N/f$ the inverse cascade persists (see for example \cite{bartello_95, Metais96, Smith02, Marino13}), although the inverse cascade becomes slower (less efficient) as $N/f$ is increased.

Finally, and more importantly, we find that the purely stratified case  is different from the other two cases (dashed, blue
line in Fig. \ref{fig:flux}). The isotropic flux $\Pi_T(k)$ at $k<k_F$ is almost zero, indicating that almost no net energy goes across spheres in Fourier space for small wavenumbers. This stems from different behaviors in the parallel and perpendicular directions. Indeed, the perpendicular flux $\Pi_T(k_\perp)$ shows a range with negative values (albeit much smaller than in the rotating case or in the rotating and stratified case), and becomes negligible for $k_\perp \lesssim 4$. The parallel flux $\Pi_T(k_\parallel)$ is positive and dominant for all wavenumbers, indicating a strong transfer towards smaller vertical scales, although this flux decreases rather sharply as we approach the largest scale.

Considering now  scales smaller than the forcing, or for $k>k_F$, we note the stronger direct (positive) flux in the purely stratified case, dominated by the transfer in the vertical direction which corresponds to the  slow modes of the system with $k_\perp=0$, whereas the transfer in the horizontal plane is stronger for pure rotation (again, for the slow mode, with now $k_\parallel=0$). 

\section{Discussion and conclusion}

In Figure \ref{fig:diagram} we present  a schematic representation of these fluxes for all cases considered (rotating, rotating and stratified, and purely stratified). The isotropic flux $\Pi_T(k)$ measures how much energy goes per unit of time across circles in the $(k_\parallel,k_\perp)$ plane (spheres in 3D Fourier space); for rotating flows, as well as for rotating and stratified flows, it is negative for $k<k_F$.  The fluxes $\Pi_T(k_\perp)$ and $\Pi_T(k_\parallel)$ measure how much energy per unit of time crosses lines respectively with constant $k_\perp$ or $k_\parallel$ (i.e., cylinders or planes in 3D Fourier space). In all the runs with non-zero rotation, all the fluxes are negative resulting in an accumulation of a fraction of the injected energy at the smallest available wavenumbers.

The purely stratified run has, for $k<k_F$, almost no net energy going across circles. This is the result of  a competition between an inverse flux in $\Pi_T(k_\perp)$ and a direct flux in $\Pi_T(k_\parallel)$. A fraction of the energy forced in the shell $k=k_F$ goes towards wavenumbers with smaller $k_\perp$, larger $k_\parallel$, or strata.
 This process is known to result in the development of vertically sheared horizontal winds \cite{Smith02}, and with sufficient scale separation as is the case in our simulations, also results in the $\sim k_\perp^{-5/3}$ spectrum observed for a range of wavenumbers with $k_\perp<k_F$. However, energy is also transferred directly towards larger $k_\parallel$, as indicated by $\Pi_T(k_\parallel)>0$, and this prevents energy from reaching $k=1$, and results in a flat spectrum for the smallest wavenumbers.
 
{Several other forcing mechanisms have been put forward in the literature, such as 2D forcing \cite{smith_96, deusebio_14}, as well as an assembly of vortex dipoles \cite{augier_13}. Each deserves separate studies in view of the complexity of these flows, and of the several characteristic scales in the system, as well as the different inertial ranges that have to be resolved a priori. The question as to whether or not universal results will ensue is still open and will require ample further studies at high resolution. We note that the aspect ratio of the fluid is also a factor to be concerned with \cite{smith_96}, as emphasized recently for rotating flows \cite{deusebio_14} as well as for stratified flows \cite{sozza_14}.}

{We therefore cannot rule out at this point a dependency 
of the observed fluxes on the specific kind of forcing used and on the strength 
of waves. Further investigations along these lines are left for a future study
where the level of anisotropy in the forcing will be varied to explore its effect 
in the cascades}.

{More complex stratified and rotating flows can also be envisaged in the future, with a wealth of new phenomena occurring, taking into consideration for example the effect of the walls on the energy budget \cite{zonta_12}, or the dependence of viscosity and diffusivity on temperature \cite{zonta_13}, leading in that case to intermittent bursts of heat transfer (see also \cite{cecilia_14} for strong intermittency in stratified flows). This opens the question of the validity of the assumption of a unit Prandtl number, at least in modeling approaches \cite{zonta_13}, the effect on thermal expansion seemingly being more important than that on viscosity \cite{zonta_14}.}

In conclusion, we have presented several direct numerical simulations of stratified and/or rotating  turbulence, with spatial resolutions up to $1024^3$ grid points, and forced at small scales to have sufficient scale separation to study the development of inverse cascades. While purely rotating flows, and rotating and stratified flows at moderate values of $N/f$ develop inverse cascades, purely stratified flows have almost zero isotropic energy flux at large scales. However, there is a small negative perpendicular flux towards small $k_\perp$ that results in the development of vertically sheared horizontal winds. Unlike in previous works, the present study shows that when sufficient scale separation is allowed between the forcing scale and the domain size, the kinetic energy spectrum at large scales displays a power law behavior in the perpendicular direction compatible with $\sim k_\perp^{-5/3}$. This spectrum is the result of the combined inverse transfer in the perpendicular direction, and a direct cascade of the large-scale horizontal winds manifested by a positive flux of energy in the parallel direction at the largest scales. This highly anisotropic transfer in stratified flows can result in the build-up of a power law spectrum at large scales (albeit up to some wavenumber, and followed at larger scales by a flat spectrum), with positive flux in some directions in spectral space.

\begin{acknowledgments}
This work was supported by NSF/CMG grant 1025183; it was also sponsored by an NSF cooperative agreement through the University Corporation for Atmospheric Research on behalf of the National Center for Atmospheric Research (NCAR). Computer time was provided on NSF/XSEDE TG-PHY100029 and 110044, and NCAR/ASD on Yellowstone. PDM acknowledges support from UBACYT Grant No. 20020110200359, PICT Grants No. 2011-1529 and No. 2011-1626, and PIP Grant No. 11220090100825. AP acknowledges support from LASP and in particular Bob Ergun. RM acknowledges the Regional Operative Program (ROP) Calabria ESF 2007/2013 and the Marie Curie Project FP7 PIRSES-2010-269297 ``Turboplasmas''.

\end{acknowledgments}

\bibliography{ap}
\end{document}